\documentclass[preprint,12pt]{elsarticle}

 \usepackage[english]{babel}
\usepackage{mathtools}
\usepackage{amsmath}
\usepackage{graphics}
\usepackage{graphicx}
\usepackage{epsfig}
\usepackage{amssymb}
\usepackage{color}

\graphicspath{{eps_figs/}}

\begin{document}
\begin{frontmatter}

\title{A new anti-neutrino detection technique  based on positronium tagging with plastic scintillators}
\def\A{\kern+.6ex\lower.42ex\hbox{$\scriptstyle \iota$}\kern-1.20ex a}
\def\E{\kern+.5ex\lower.42ex\hbox{$\scriptstyle \iota$}\kern-1.10ex e}
\author[mi]{G.Consolati}
\author[apc]{D.Franco}
\author[iphc]{C.Jollet}
\author[iphc]{A.Meregaglia}
\author[iphc]{A.Minotti}
\author[apc]{S.Perasso}
\author[apc]{A.Tonazzo}

\address[mi]{Department of Aerospace Science and Technology, Politecnico di Milano, via La Masa 34, 20156 Milano, Italy}
\address[apc]{APC, Univ. Paris Diderot, CNRS/IN2P3, CEA/Irfu, Obs. de Paris, Sorbonne Paris Cit\'e, 75205 Paris, France}
\address[iphc]{IPHC, Universit\'e de Strasbourg, CNRS/IN2P3, 67037 Strasbourg, France}


\begin{abstract}

The main signature for anti-neutrino detection in reactor and geo-neutrino experiments based on scintillators is provided by the space-time coincidence of positron and neutron produced in the Inverse Beta Decay reaction. Such a signature strongly suppresses backgrounds and allows for measurements performed underground with a relatively high signal-to-background ratio. In an aboveground environment, however, the twofold coincidence technique is not sufficient to efficiently reject the high background rate induced by cosmogenic events. Enhancing the positron-neutron twofold coincidence efficiency may pave the way to future aboveground detectors for reactor monitoring. 
We propose a new detection scheme based on a  threefold coincidence, between the positron ionization, the ortho-positronium (o-Ps) decay, and the neutron capture, in a sandwich detector with alternated layers of plastic scintillator and aerogel powder. We present the results of a set of dedicated measurements on the achievable light yield and on the o-Ps formation and lifetime. The efficiencies for signal detection and background rejection of a preliminary detector design are also discussed.
\end{abstract}

\end{frontmatter}

\section{Introduction}

The Inverse Beta Decay reaction (IBD), $\bar{\nu}_e + p \to e^+ + n$, is most commonly used to detect electron anti-neutrinos, thanks to the coincidence between the emitted positron and the neutron capture. In scintillators, positron ionization  occurs in a time window of hundreds of picoseconds from the IBD reaction, providing the first signal in time. The second signal is due to the gammas from  neutron captures, delayed by a few up to  tens of microseconds with respect to the IBD start time.  The coincidence technique between positron and neutron is generally improved by looking at their reconstructed positions and by imposing a space correlation.

This technique was successfully exploited by Cowan and Reines in 1956  for the first neutrino detection. The positron-neutron coincidence  represents also the main signature in the Double Chooz~\cite{Abe:2011fz}, DayaBay~\cite{An:2012eh} and RENO~\cite{Ahn:2012nd} experiments, aiming to measure the $\theta_{13}$ neutrino mixing angle. In these experiments, accidental and correlated backgrounds are kept under control by installing the detectors at tens of meters underground and equipping them with active and passive shielding. The detection of anti-neutrinos from beta decays of nuclides naturally occuring in the Earth (geo-neutrinos) \cite{Bellini:2013nah,Araki:2005} is also based on the same technique.
 
Currently, several experiments are designed to observe anti-neutrinos with detectors located at shallow depths or aboveground, and at a few (tens of) meters  away from  reactor cores for  monitoring the reactor power (e.g. PANDA~\cite{Kuroda:2012dw} and NUCIFER~\cite{Porta:2010zz}), and/or  for testing the sterile neutrino hypothesis (e.g. DANSS~\cite{Belov:2013qwa} and STEREO~\cite{stereo}). However, because of the high rate of gammas and neutrons emitted from the reactor facilities and from  cosmic rays, the twofold coincidence technique alone is not sufficient to reach a favourable signal-to-background ratio. The typical solutions adopted to improve the signature are detector segmentation \cite{Belov:2013qwa}, which allows to better identify the topology of the positron annihilation gammas, and pulse shape discrimination \cite{Porta:2010zz}, which provides a separation between neutron-induced proton recoils and electron-like events.

In this work, the IBD signature is enhanced by looking at the decay of ortho-positronium (o-Ps), a process competiting with positron annihilation. After ionisation, positrons can form o-Ps states, which, in matter, can reach a lifetime up to tens of nanoseconds. The positron ionization component and the gammas from the o-Ps decay can then be separated in time in order to form, in association with the neutron induced signal, a threefold coincidence.
 
The o-Ps formation probability and lifetime have been already measured in the most commonly used liquid scintillators~\cite{Kino:2000,Franco:2010rs,Consolati:2013rka} to be about 50\% and 3~ns respectively. 
Such a decay time is comparable to the fast scintillation component of liquid scintillators, thus the identification of o-Ps decays is difficult. Nonetheless, the Borexino collaboration~\cite{Collaboration:2011nga} exploited the scintillation pulse shape distortion induced by the o-Ps decays to identify and reject cosmogenic $^{11}$C $\beta^+$ decays, the dominant  background in the solar {\it pep} neutrino rate measurement. The Double Chooz experiment has also exploited this technique to select anti-neutrino candidates~\cite{Abe:2014uba}, even though with a very poor efficiency.

\begin{figure}[t]
\begin{center}
\includegraphics[width=7cm]{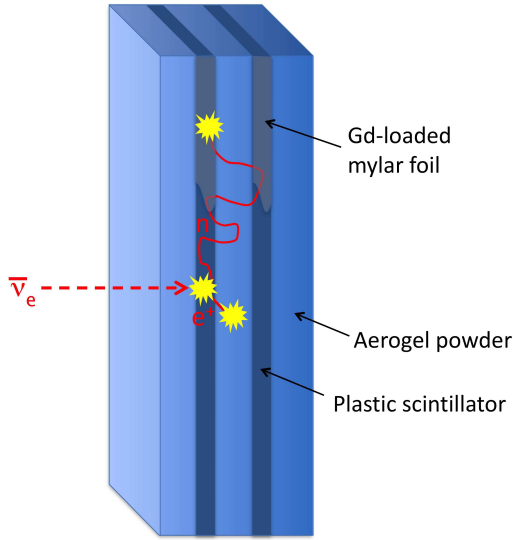}
\caption{Schematic view of the proposed detector concept. Anti-neutrinos interact in the active layers, made of plastic scintillators, and the positrons annihilate in the passive layers, made of aerogel. The delayed neutron capture signal is provided by the Gd doping in the reflecting Mylar$^{\textregistered}$ wrapping the scintillator bars.}
\label{fig:detdesign}
\end{center}
\end{figure}

The large reduction of o-Ps lifetime in scintillators, from the mean value of 142 ns in vacuum, is mainly due to the pick-off quencher process, which occurs when o-Ps collides with a closed-shell atom. The larger the cavity in which o-Ps gets trapped, the smaller the collision probability, therefore, in porous materials, the pick-off effect is strongly suppressed. This paper proposes a solution by defining a segmented detector, alternating porous material (aerogel)  and fast plastic scintillator layers, optimized to detect anti-neutrinos. The threefold coincidence technique is capable to identify the positrons from anti-neutrinos, originated inside a plastic layer and  escaping  into the porous material where they may form the o-Ps state. The first signal of the triple coincidence is represented by the scintillation pulse generated by the positron ionization and the second  one by the o-Ps decay gammas interacting within the plastic layers. The third coincidence is due to neutron capture on Gadolinium, present in the coating material of the plastic layers, as shown in figure~\ref{fig:detdesign}. 

This paper aims at investigating the efficiency of the threefold coincidence and its background rejection power, by first characterizing the porous and the active materials, and then by simulating an ideal geometry.

\section{Measurement of the o-Ps properties in Aerogel}
\label{sec:ops}

\begin{table}[t]
\begin{center}
\begin{tabular}{lcc}
\hline
& o-Ps & o-Ps   \\
Material  &  fraction  &  lifetime \\
  &  [\%] &  [ns] \\

\hline
Standard  Cabot$^{\textregistered}$ Aerogel    & $29.6 \pm 1.9$ & $58.8 \pm 0.7$\\
Lumira$^{\textregistered}$   Aerogel  Particles & $25.7 \pm 2.6$ & $60.2 \pm 2.6$\\
Pyrogel$^{\textregistered}$ XT Blanket & $8.6 \pm 0.8$ & $52.4 \pm 2.0$\\
Spaceloft$^{\textregistered}$ Blanket  & $14.4 \pm 1.3$ & $53.8 \pm 1.6$\\
Cryogel$^{\textregistered}$ Z Blanket  & $19.0 \pm 1.5$ & $51.9 \pm 0.8$\\
Airloy$^{\textregistered}$  X103 Strong Aerogel 200 mg/cc & $9.1 \pm 0.7$ & $47.3 \pm 1.0$\\
Airloy$^{\textregistered}$   X103 Strong Aerogel 400 mg/cc & $6.5 \pm 0.5$ & $47.9 \pm 1.1$\\
\hline
\end{tabular}
\caption{o-Ps formation fraction and lifetime for the different arogel-based materials.}
\label{tab:aerogel}
\end{center}
\end{table}%

The identification of porous material with the longest o-Ps lifetime and highest formation probability was achieved by characterizing different classes of materials, such as  4--10 nm porous glasses, nanoporous silica based materials, silica based aerogel, and syndiotactic polystyrene. Plastic scintillators doped with silica nano-particles were also investigated, but a degradation of the optical properties of the scintillators was observed. The class of materials yielding the best characteristics in terms of o-Ps properties is \textit{aerogel}. 

\begin{figure}[t]
\begin{center}
\includegraphics[width=0.6\linewidth]{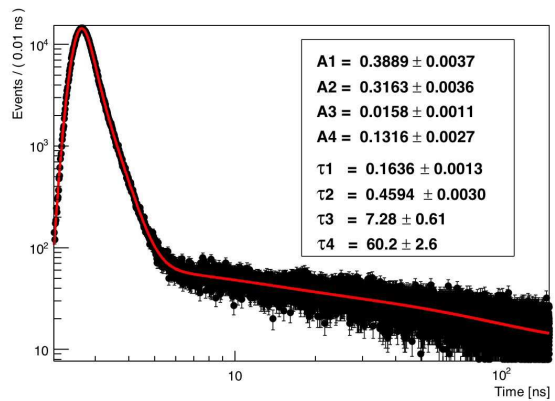}
\caption{Time distribution of the positron annihilation and o-Ps formation in Lumira$^{\textregistered}$ aerogel powder with a fit made of four exponentials and a constant convoluted with a gaussian error.}
\label{fig:fitpowder}
\end{center}
\end{figure}

Aerogel is a class of synthetic porous ultralight materials obtained by replacing the liquid component of a gel with a gas. The result is an extremely low density solid with high porosity, where the pores size depends on the chemical compound and on the production process. In particular, we  characterized  seven commercial aerogel based materials (listed in table~\ref{tab:aerogel}), in air environment and at room temperature, with a standard Positron Annihilation Lifetime Spectroscopy (PALS) technique. A $^{22}$Na source was deployed inside the aerogel powder or between two aerogel bulk samples, so that the positron can penetrate inside the sample itself. A BaF$_2$ based detector, used as a trigger, detects the 1.273 MeV gamma emitted in the $^{22}$Na decay, and a second identical detector measures the delay, with respect to the trigger time, of the  positronium decay gammas. The efficiency of the setup was evaluated with Monte Carlo simulations. The spectral time distribution is fitted with a four exponential distribution to take into account the annihilation in the titanium supporting structure and in the aerogel, and the para- and ortho-positronium components. A  fit of the distribution is shown in figure~\ref{fig:fitpowder}. For more details about the setup  and the data analysis see reference~\cite{Franco:2010rs} and \cite{Consolati:2013rka}.

\label{sec:aerogel}

The longest measured o-Ps lifetimes and highest formation probabilities were found for  the Lumira$^{\textregistered}$ powder and the Hydrophobic Silica Disc (Standard Cabot$^{\textregistered}$ Aerogel), as quoted in table~\ref{tab:aerogel}. The two mean lives are $\sim$60~ns, while in powder o-Ps  has a lower formation probability ($\sim$26\%) with respect to the bulk ($\sim$30\%). The latter, however, has an extremely fragile structure which hinders the production and the handling of large aerogel panels. Lumira$^{\textregistered}$ powder aerogel, despite its slightly lower formation probability,  is the chosen porous material candidate, since it can be easily  produced in large quantity, with reduced  costs,   and, being a powder, it can fill any volume without additional manufacturing.

\begin{figure}[t]
\begin{center}
\includegraphics[width=0.6\linewidth]{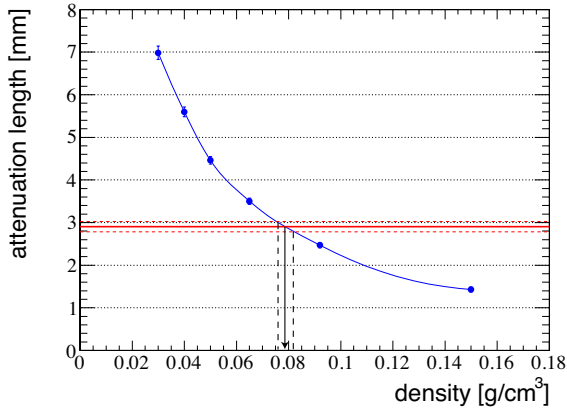}
\caption{Positron attenuation length in aerogel for different densities for MC (blue line) and for the data (red line). The dashed red lines correspond to the $\pm 1~\sigma$ region of the value computed for the data. The vertical black arrow correspond to the density giving the best agreement between data and MC whereas the black dashed lines correspond to the $\pm 1~\sigma$ region.}
\label{fig:density}
\end{center}
\end{figure}

The effective density of the powder, critical for correctly assessing  the positron thermalization in Monte Carlo simulations, has been determined  by measuring the o-Ps fraction by varying the  aerogel powder sample thickness, around the $^{22}$Na source,  from 1 to 4~cm.  The o-Ps component is directly related to the positron attenuation length in aerogel, which resulted in $2.90 \pm 0.12$~mm.   Comparing this value with  simulated ones with different aerogel densities, we found  the best agreement   for a density of $0.079 \pm 0.003$ g/cm$^3$, as shown in figure~\ref{fig:density}. This result is consistent with the nominal bulk aerogel density of about 0.1  g/cm$^3$.

\section{Detector configuration}
\label{sec:detector}

Plastic scintillator layers have the twofold role of thermalizing the positron by ionization and of detecting gammas from the o-Ps decay. The detection signature requires that the positron, once thermalized, escapes the plastic layer to reach the porous material. Plastic layers have to be then sufficiently thick to fully thermalize the positron, but sufficiently  thin for avoiding its trapping. The plastic layer thickness has to be comparable with a few positron attenuation lengths, nominally from a few millimeters to about one centimeter. Such a thickness, however, is not suitable for the detection of gammas from o-Ps decay ($\sim$0.5~MeV), whose mean interaction lengths is about 10~cm. In order to increase the gamma detection efficiency, the introduction of layers of thick (10~cm) plastic scintillators inside and around the detector is also considered. 

\begin{figure}[t]
\begin{center}
\includegraphics[width=11cm]{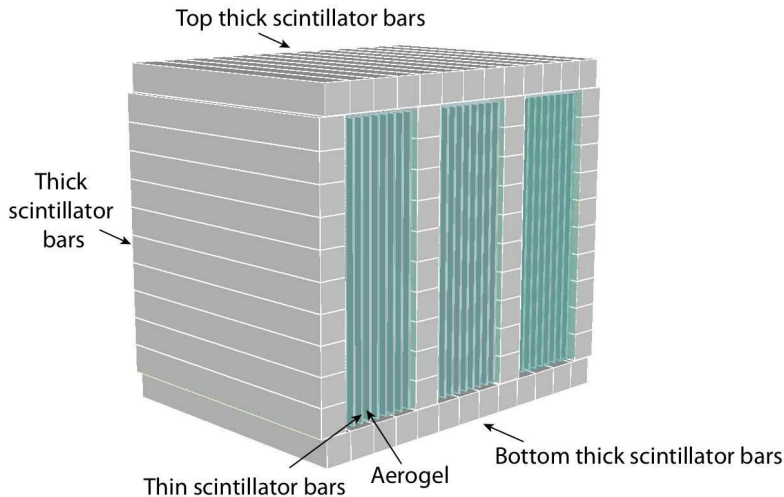}
\caption{Schematic view of the proposed detector layout.}
\label{fig:detector1}
\end{center}
\end{figure}

The detector  optimization was carried out by simulating several geometries with Geant4~\cite{geant4, geant4_1}, and, in particular, by varying  the scintillator and aerogel layer thicknesses in order to maximize the number of detected neutrinos. The reference configuration for the optimization is  1~m$^3$ total detector volume, assuming a location  at 20~m from a 4.5~GW reactor core, which corresponds  to $\sim$7.6 interacting neutrinos per day and per kilogram of plastic scintillator. Each layer  has a surface of about $100 \times 97.5$~cm$^2$, and a thickness varying from 10 to 45 mm  for the aerogel, and from 5 to 15 mm for the plastic. Each plastic layer is made of 39 bars 1 meter long and 2.5~cm wide, and each bar is wrapped with Gd-coated reflecting mylar foils. The wrapping has the double function of capturing neutrons, thanks to the very high Gd cross section, and of  reflecting the light  throughout the bar up to the extremities, where it is collected. A scheme of the detector is shown in figure~\ref{fig:detector1}.

 In total, we tested 350 detector configurations. The number  of detected anti-neutrinos (N$_{\nu}$) is defined as the number of  IBD (N$_{IBD}$) reactions occurring in the scintillator layers multiplied by five factors: 
\begin{equation}
N_{\nu} = N_{IBD} \times f_{e^+} \times f_{oPs} \times f_{\gamma,oPs} \times f_{n}\times f_{\gamma,n} 
\end{equation}
where
\begin{itemize}
\item $f_{e^+}$ is the fraction of positrons producing a signal in the scintillators, escaping the scintillator and stopping in the aerogel material;
\item $f_{oPs}$ is the fraction of positrons  forming an o-Ps state in aerogel;
\item $f_{\gamma,oPs}$ is the fraction of o-Ps decay gammas interacting with the scintillator bars;
\item  $f_{n}$ is  the fraction of  neutrons captured on Gd;
\item  $f_{\gamma,n}$ is  the fraction of gammas from IBD neutron captures on Gd producing a signal in the scintillator. 
\end{itemize}
We fixed $f_{oPs}$ to 26\% from the measurements described in the previous section, whereas the other four factors resulted from the Geant4 simulations for each detector configuration. To correctly simulate the IBD events and the detector response with respect to o-Ps,  we simulated the IBD  positron and neutron,  in the same vertex. The anti-neutrino energy spectrum is computed according to reference~\cite{Schreckenbach:1985ep,VonFeilitzsch:1982jw,Hahn:1989zr} whereas the positron and neutron kinematics is obtained from the GENIE neutrino interaction generator~\cite{Andreopoulos:2009rq}. In addition we wrote ad-hoc classes in Geant4 to simulate the positronium formation and decay. At this stage, each of the three expected signals from the three-fold coincidence (positron ionization,  positronium decay and neutron gamma interactions) is required to  deposit at least 200 keV in  scintillator. 

The detector configuration which maximizes the number of detected neutrinos consists of 27 layers of 10~mm of plastic scintillators interleaved with 22.5~mm thickness layers of aerogel. In addition four thick layers of plastic scintillator surround the detector and two more  are inserted  inside as shown in figure~\ref{fig:detector1}. This configuration  corresponds to a total active mass of 263 kg of thin plastic bars.

\section{Optical modules}

The scintillator is required to have a fast rise time and a  short decay time of the fast component to enhance the o-Ps detection. The chosen candidate, Eljen EJ-200~\cite{eljenbar}, fulfills such requirements with a rise time of 0.9 ns and a decay time of 2.1 ns. 

\begin{figure} [tbp]
\begin{center}
\includegraphics[width=0.6\linewidth]{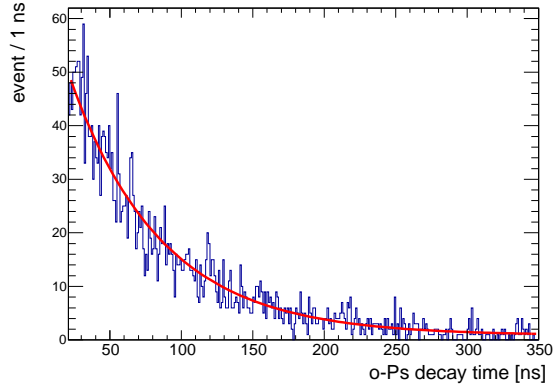}
\caption{Time difference between the 1.27~MeV gamma emitted in the $^{22}$Na, and 511~keV gamma issued by the o-Ps decays. The red line represents a fit with two exponentials: one corresponding to the o-Ps decays and the other one to accidental coincidences.}
\label{fig:oPssinglebar}
\end{center}
\end{figure}

A dedicated measurement was conceived to prove the scintillator capability in  separating in time   the positron ionization signal and the light pulse  induced by the decay gammas. A vial containing aerogel powder and a low activity (500 Bq) $^{22}$Na source was located next to a scintillator bar coupled with two PMTs (the measured light yield in this configuration is $\sim$342  p.e./MeV). $^{22}$Na emits in coincidence a 1.27~MeV gamma and a positron. The 1.27~MeV gamma was observed as the prompt signal in the scintillator, mimicking the positron ionization signal expected from the anti-neutrino interaction. The decay of the positronium state, formed by the positrons in aerogel, provided the delayed signal in the same scintillator bar. Looking at the coincidence between the two signals in a time window between 20 and 450~ns, we measured  an o-Ps lifetime of $64 \pm 2$~ns, in  agreement with the previously measured value of $60 \pm 3$~ns, as shown in figure~\ref{fig:oPssinglebar}. The fraction of accidental events is consistent with the expectation of  0.7\%. 

The large number of scintillator bars foreseen for the optimal detector configuration (1053 thin and 66 thick) does not allow to directly couple photomultiplier tubes to the extremities of each of them. We therefore considered the possibility of clustering bars together into optical modules, each module coupled to a pair of PMT's. The clustering increases the probability to observe the two positron induced signals, prompt and delayed, in the same module. The previous measurement demonstrates the capability to separate the two signals when they occur with a delay greater than 20 ns. Interactions into two different modules allows to look for delayed coincidences down to a few nanoseconds. 

\begin{figure}[t]
\begin{center}
\includegraphics[width=0.6\linewidth]{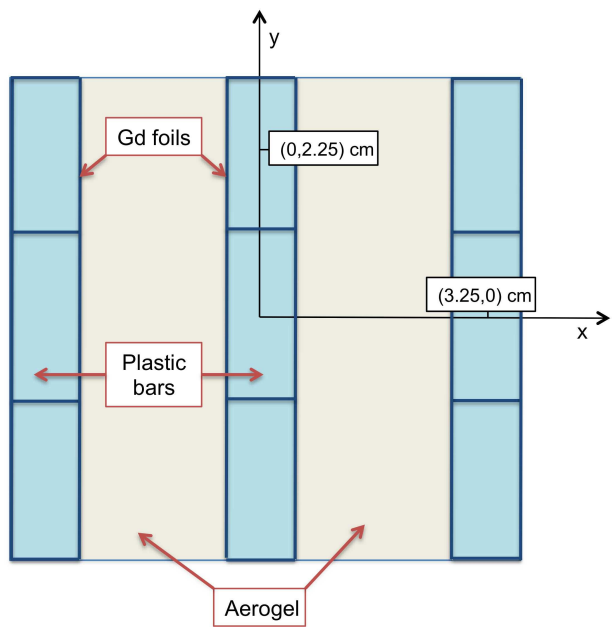}
\caption{Scheme of the optical module xy surface. }
\label{fig:module}
\end{center}
\end{figure}

The clustering, however, affects the light yield. We investigated three options for clustering bars together in optical modules in order to minimize the loss of scintillation light.  At first, we exploited optical fibers coupled to multi-anode PMTs. In particular, we tested two configurations: St. Gobain BCF-92MC fast wavelength shifting fibers  glued in a specially machined groove along the bar, and clear quartz fibers  inserted at the end of special light guides coupled to the plastic bars. In both cases, the measured light yield was about a few tens of photoelectrons per MeV, too low for the needs of the here proposed detector. The most efficient solution is provided by a light guide coupled to 9 scintillator bars, as shown in figure \ref{fig:module}. The light guide produced by Eljen~\cite{eljenbar} is coupled to a 3--inch 9821B Electron Tube PMT~\cite{electrontube}, characterized by a  fast single photoelectron rise time of 2.1 ns.  The light guide is a square frustum 5.1 cm high, where the large surface base (7.5$\times$7.5 cm$^2$) is designed to entirely cover the 9 scintillator bars and the aerogel regions, and the small base (5.3$\times$5.3 cm$^2$) to match the flat area of the 3--inch PMT.  

To measure the light yield, a  EJ-200 scintillating bar of 10$\times$25$\times$1000~mm$^3$, coupled  with  light guides, in turn coupled to two PMTs, was exposed to  an electron spectrometer  emitting 1.8 MeV electrons.  The electron beam was focused to interact  in the middle of  the bar. The portion of the light guides not coupled to the scintillator bar was covered by a reflecting material.   The measurement was repeated by moving the bar in different positions with respect to the light guide center, as quoted in table \ref{tab:lightyield}.  We observed a maximum light yield of 272 $\pm$ 1 (stat.) p.e./MeV when the bar  is in contact with the center of the light guide base. Moving the bar face towards the edges of the light guide, the light yield degrades by  a maximum of  16\%, as reported in table \ref{tab:lightyield}. Such a light yield guarantees the detector resolution to well identify the anti-neutrino energy spectrum.  

\begin{figure}[htbp]
\begin{center}
\includegraphics[width=0.6\linewidth]{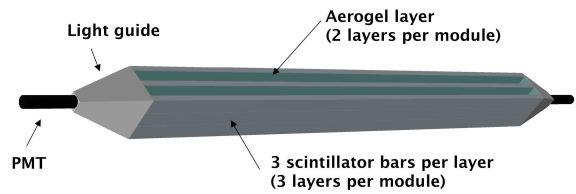}
\caption{Module of 9 scintillating bars read by a single PMT on each side.}
\label{fig:module2}
\end{center}
\end{figure}

\begin{figure}[htbp]
\begin{center}
\includegraphics[width=0.6\linewidth]{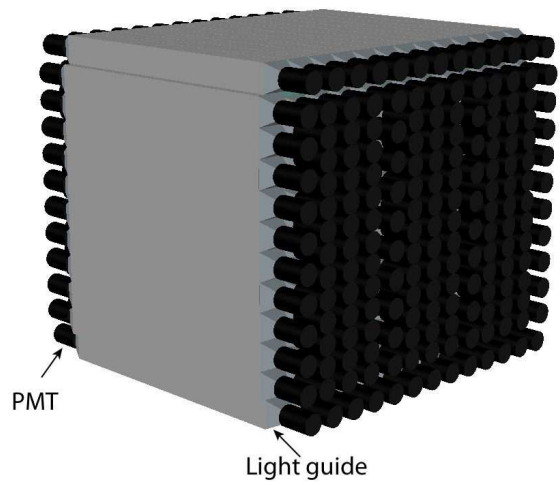}
\caption{Schematic view of the proposed detector layout with  light guides and  PMTs.}
\label{fig:detector}
\end{center}
\end{figure}

In its final configuration, the detector consists in 117 optical modules, each consisting of 9 scintillator bars poured in aerogel powder, as shown in figures \ref{fig:module} and \ref{fig:module2}.  A thickness of 22.5 mm of aerogel  separates two modules along the x-axis (see figure \ref{fig:module}). A view of the detector is shown in figure \ref{fig:detector}.

\begin{table}[t]
\begin{center}
\begin{tabular}{llc}
\hline
Position & x, y  &  Light Yield    \\
& [cm]&  [p.e./MeV]    \\
\hline
Center & 0, 0 &   272  \\
Top/Bottom & 0, $\pm$2.25  & 228 \\
Corner & $\pm$3.25, $\pm$2.25  & 232  \\
Side & $\pm$3.25, 0   & 252  \\
\hline
\end{tabular}
\caption{Light yield at the center of the bar, for different bar positions inside the cluster module.}
\label{tab:lightyield}
\end{center}
\end{table}%

\section{Signal detection and background rejection efficiencies}

The expected main background components,  limiting the neutrino detection, are  external gammas and cosmogenic and reactor-induced neutrons. Gammas are originated from natural radioactivity inside the detector materials and from the environment. They generate only accidental background, since a single gamma can not reproduce the  triple  coincidence, defined by the positron ionization signal, the positronium decay, and the neutron capture. Selection cuts  based on time coincidences only provide high efficiency for rejecting gammas. Further requirements on the energy windows and on the topology of the events can suppress this background component to negligible levels. On the other hand, neutrons are extremely dangerous since they produce a correlated background, potentially mimicking the expected neutrino signal.  

A full Monte Carlo simulation of the optimal detector configuration was developed for evaluating the signal detection and the background rejection efficiencies. The simulation includes full photon tracking, taking into account the optical properties of the light guides, the scintillator bars, and the PMT cathodes. The Electron Tube 9821B PMT quantum efficiency was derived from the spreadsheet of the manufacturer \cite{electrontube}. The single electron rise time and jitter of the  PMTs  are equal to 2.1~ns and 2.2~ns respectively. The scintillation photon yield was tuned  in order to match the measured light yield of 272  photoelectrons/MeV, at the center of the scintillator bar.   The characteristic  fast and slow scintillation decay times are equal to 2.1~ns and 14.2~ns, respectively, with a fast to slow component ratio of 0.73. The scintillation rise time of 0.9~ns and the index of refraction of 1.58 were also included  as provided by the manufacturer~\cite{eljenbar}.  
We included the proton quenching from reference~\cite{quench}. The mean neutron capture time on the Gd foils, coating the scintillation bars, was found to be $\sim$62~$\mu$s. 

An ad hoc reconstruction algorithm was developed for identifying  light pulses in each bar. A pulse is defined if a minimum of 10 photons are detected  in a time window of 20 ns. The integration time window is maximized up to the beginning of a second pulse, if any, in the same module. The maximum integration window is 100 ns.   A second pulse, in the same module, is identified if the time interval from the end of the previous pulse is at least 20 ns, much larger than $\sim$5~ns, the maximum time needed by photons to cover 1~m of scintillator, without reflection and scattering. The starting time of the pulse is defined at  30\% of the peak value of the time distribution, once subtracted the time of flight of the photons.

 We did not consider at this level the possibility to exploit the pulse shape discrimination, which represents an extra  efficient technique to reject neutron background, since we did not include in the simulation the proper scintillation singlet to triplet ratios, depending on the nature of the interacting particle.

To evaluate the neutrino detection and the neutron rejection efficiencies, we simulated 10$^5$ IBD interactions in the scintillator bars, and 30 samples of 10$^5$ monoenergetic neutrons in an energy range from 1~MeV to 10~GeV. 
On these samples, we tuned the selection cuts in order to maximize both  the signal efficiency and background rejection power.

At this stage, we consider fake signals induced by single neutrons only. Other sources of background, like multiple neutrons induced by a single cosmic muon shower,  were not part of this study but potentially might represent a  serious background despite the triple coincidence.


The first requirement in the selection concerns the thick plastic scintillator bars, conceived for both maximizing the gamma detection and for vetoing the external background. In this view, any event with the first pulse in the external thick bars is discarded. 
The ionization positron signal (the first pulse) is required to have an energy lower than 1500 p.e. The positronium decay signal is searched  in a time window of 300~ns from the prompt pulse.  The gammas arising from the o-Ps decay are required to provide at least 2 signals in two separate modules in a time interval of less than 5~ns.  At least one pulse occurring after 300~ns, which is assumed to belong to gammas from neutron captures, is required.  

\begin{figure}[t]
\begin{center}
\includegraphics[width=0.7\linewidth]{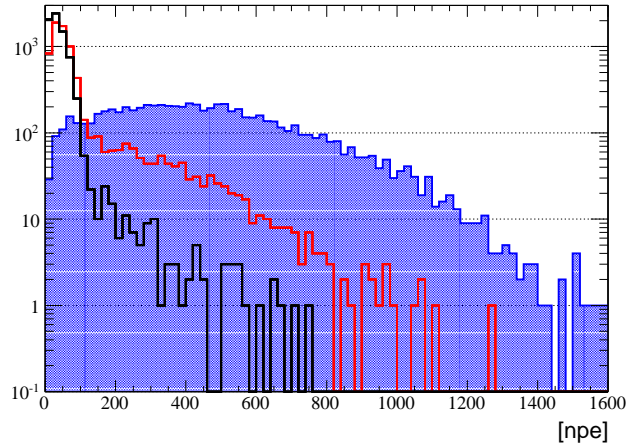}
\caption{Number of photoelectrons, from the neutrino simulations, of the first pulse (filled blue), corresponding to the positron ionization, and of the second (red) and third (black) pulses due to the positronium decay gammas.}
\label{fig:spectra}
\end{center}
\end{figure} 

The energy spectra, given in number of photoelectrons, of the first three pulses are shown in figure \ref{fig:spectra}. The energy spectrum of the first pulse reproduces that expected from the neutrino-induced positron. The second and the third pulse spectra are mostly related to the Compton scattering of the 511~keV gammas. The high energy component of these spectra, above the maximum Compton energy at $\sim 340$~keV, is due to the overlap, in the same pulse, of multiple gammas from positronium decay or from neutron capture. 

\begin{figure}
\begin{center}
\includegraphics[width=0.7\linewidth]{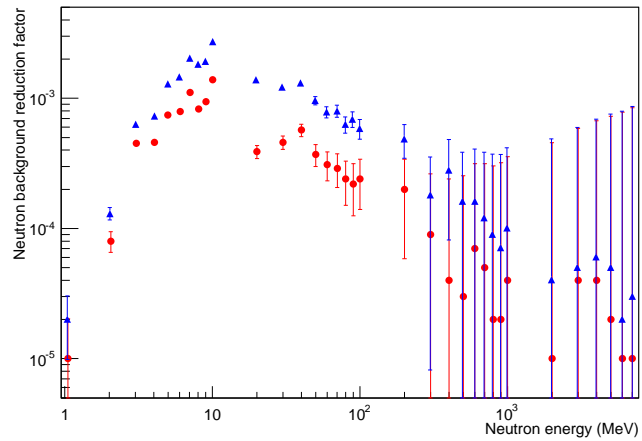}
\caption{Neutron reduction factor  for the o-Ps decay search in the  [5, 300] ~ns  (blue triangles) and  [10, 300]~ns (red circles) time windows.}
\label{fig:neutrons}
\end{center}
\end{figure}

The time delay between the prompt ionization positron signal and the first positronium decay gamma interaction can be opportunely tuned in order to maximize the signal or better suppress the neutron background. Increasing  this delay time better constrains the positronium sample while limiting the selection efficiency. At the same time, it reduces the possibility of neutron contamination. For instance, a delay time cut at 10~ns provides a selection efficiency of 1.9\% with a minimum neutron suppression factor of $\sim 1.5 \times 10^{-3}$ at a neutron energy of $\sim$10~MeV, as shown in figure \ref{fig:neutrons}. The rejection power increases by 1--2 orders of magnitude at higher energies. 

Reducing the delay time cut to 5~ns, the signal efficiency can be boosted recovering a fraction of the positronium events formed inside the plastic bars. Positrons stopping inside the plastic scintillator  are about 3 times larger with respect to those stopping in aerogel layers. The o-Ps formation fraction and lifetime in plastic are $\sim 40\%$ and  $\sim$2.2~ns respectively. Setting the delay time cut at 5~ns for the time window of the o-Ps decay, some events can still be measured, reaching a signal efficiency of $\sim$5.1\%. At the same time, the neutron rejection power is reduced, with respect to the 10~ns cut, by a factor $\sim$2 (see figure~\ref{fig:neutrons}). A summary list of the section cuts can be found in table~\ref{tab:cuts}.

Assuming a detector located at 20~m from a 4.5~GW reactor core, the number of IBD's occurring in the plastic scintillator layers is $\sim$1920 events per day in 263 kg of active mass. The number of detected anti-neutrinos per day is expected to be between 36 and 100, depending on the selection cuts. The mass of the detector can be tuned on the basis of the envisaged physics goal. The limited overall detection rate per unit of mass ($\sim$140--380 events per day per ton at 20~m from a 4.5~GW reactor), however, is not suited for applications with research reactors with a power in the sub-GW range.

\begin{table}
\begin{center}
\begin{tabular}{lll}
\hline
\# & Variable & Requirement \\
\hline
1 & Each Pulse Energy & $>$ 10 p.e.\\
\hline
2 & Prompt Position & not at the detector boundary\\
3 & Prompt Energy & $<$1500 p.e.\\

\hline
4 & Second Pulse Delay & $<$300~ns\\
5 & Second Pulse Delay & $>$5~ns (10~ns)\\
6 & Second and Third Pulse Positions & not in the same module\\
\hline
7 & Neutron Pulse Delay &  [300~ns, 200$\mu$s]\\
\hline
\end{tabular}
\caption{Selection cuts list.}
\label{tab:cuts}
\end{center}
\end{table}%

\section{Conclusions}
We propose a new detection technique to observe electron anti-neutrinos based on a threefold coincidence which relies on the observation of the o-Ps produced in aerogel. We optimized the  design of a detector based on the proposed technology and the outcome is a sandwich like detector made of layers of plastic scintillator and layers of aerogel. A dedicated Monte Carlo was developed to optimize  the detector design by maximizing the neutrino interaction rate and the background rejection power. In addition measurements on the optical properties were carried out  to demonstrate the feasibility of the proposed detector.

Despite the relatively low efficiency in neutrino detection, the proposed detector  has  the potential to abate the neutron background, the most dangerous one for neutrino detection, with extremely high rejection power. The main advantage of such a detection technique is the possibility to reach a large signal to background ratio in a high background environment,  like aboveground or next to a reactor.

 The proposed design was intended as a proof of principle: the target mass and the detector dimensions can be opportunely tuned to increase the absolute rate of observed anti-neutrinos. If needed,  background can be further rejected with standard active vetoes or passive shielding  surrounding the  detector. An extra rejection power may be obtained by exploiting the gamma/proton pulse shape discrimination,  not included in this work.  Finally, improvements on the detection efficiency can be further reached selecting innovative porous materials, like low density aerogel bulk materials.

\section{Acknowledgments}
We acknowledge the financial support from the ANR NuToPs project (grant 2011-JS04-009-01) and from the UnivEarthS Labex program of Sorbonne Paris Cit\'e (ANR-10-LABX-0023 and ANR-11-IDEX-0005-02). 


\end{document}